\documentclass{gandb}
\usepackage{epsfig}
\usepackage{float}

\def\half{{\textstyle\frac{1}{2}}}
\def\oqu{{\textstyle \frac{1}{4}}}

\def\eqref#1{(\ref{#1})}
\let\<=\langle
\let\>=\rangle

\begin{document}
\title{EMITTANCE GROWTH OF DIFFERENT DISTRIBUTIONS}
{ON THE STABILITY AND EMITTANCE GROWTH\\
OF DIFFERENT PARTICLE PHASE-SPACE\\
DISTRIBUTIONS IN A LONG MAGNETIC\\
QUADRUPOLE CHANNEL}
\author{J.~STRUCKMEIER and J.~KLABUNDE}
\authorhead{J.~STRUCKMEIER, J.~KLABUNDE AND M.~REISER}
\address{\it Gesellschaft f\"ur Schwerionenforschung (GSI),
Postfach~110541,\\ 6100~Darmstadt, W.~Germany\\~\\
{\rm and}}
\author{M.~REISER}
\address{Electrical Engineering Department and Department of Physics and Astronomy, University of Maryland, College Park, Maryland 20742, U.S.A.}
\received{(Received October 17, 1983, revised October 15, 2026)}
\abstract{The behavior of K-V, waterbag, parabolic, conical and Gaussian distributions in periodic quadru\-pole channels is studied by particle simulations.
In the parameter range studied, the response associated with the K-V instabilities becomes progressively less pronounced toward the Gaussian distribution.
For the third-order instability studied here, the emittance growth factor of an initially Gaussian beam after a large number of periods is considerably lower than that of an equivalent K-V distribution.\newline\indent
In addition, the non-K-V distributions exhibit a rapid initial emittance growth within approximately one period of the channel, which becomes very significant at high beam intensities.
This growth is attributed to the homogenization of the space-charge density, resulting in a conversion of electric-field energy into transverse kinetic and potential energy.
Two simple analytical formulae based on different energy-partition assumptions are derived as estimates for this effect and are compared with the results obtained from particle simulations.
They are not rigorous lower and upper bounds for arbitrary beam evolution.
\\[6pt]
Revised version of: Particle Accelerators {\bf 15}, 47--65 (1984)}
\maketitle
\section{INTRODUCTION\label{sec1}}
Recent analytical and simulation studies by Hofmann, Laslett, Smith and Haber\cite{ref1} show that the well-known K-V (Kapchinskij-Vladimirskij) distribution\cite{ref2} is subject to instabilities in a periodic focusing channel.
The region of these instabilities can be defined in terms of the phase advance of the particle oscillation per channel period without space charge, $\sigma_0$, and with space charge, $\sigma$.
For a non-K-V distribution, $\sigma$ is used here as the phase advance of the initial rms-equivalent matched K-V beam, not as a common phase advance of all particles.
Phase advances quoted as lattice parameters are given in degrees.
They are to be expressed in radians in the dimensional formulae of Section~\ref{sec:2.3}, whereas their ratios are independent of this choice.
From these studies, the above authors concluded that one should impose the restriction $\sigma_0\le 60^\circ$ to avoid the dangerous second-order (envelope) instabilities (occurring for $\sigma_0 > 90^\circ$) as well as the third-order mode ($\sigma_0 > 60^\circ$).
In addition, they recommend to limit the beam intensity such that $\sigma\ge 24^\circ$ to avoid fourth- or higher-order instabilities.
They point out, however, that this latter restriction may be too conservative in view of the fact that simulation studies for $\sigma_0\le 60^\circ$ indicated that these higher-order modes saturate at low levels and that the rms emittance of a K-V distribution is not affected for $\sigma$ values as low as $6^\circ$.

\;\;In connection with the beam transport experiments at the University of Maryland\cite{ref3} and at GSI,\cite{ref4} we studied the behavior of K-V,
waterbag, parabolic, conical and Gaussian distributions by particle simulation with the modified PARMILA code at GSI.
Our goal was to determine whether more ``realistic'' non-stationary initial distributions are also affected by the instabilities found in a matched K-V beam, which provides an exact self-consistent reference distribution for the ideal linear focusing channel.
In a periodic channel, matched self-consistency means that the distribution repeats with the lattice in the appropriate normalized coordinates, rather than remaining independent of position in laboratory coordinates.
Each of these distributions is artificially produced according to the appropriate phase-space density function.
The family allows us to compare beams with progressively more centrally peaked spatial profiles and nonlinear self-fields.
Of the different types considered, the truncated Gaussian distribution provides a useful model of a smooth, centrally peaked laboratory beam, without being a universal model of measured distributions.

In a previous paper,\cite{ref5} we reported the results of analytical and simulation work concerning envelope oscillations and (second-order) instabilities of mismatched beams.
The particle simulation studies were restricted to K-V and Gaussian distributions and showed that, for the cases considered ($\sigma_0 > 90^\circ$), the effect of instabilities on an initial Gaussian distribution is even worse than for a K-V beam.
These findings support the recommendation of Ref.~1 to avoid the region $\sigma_0 > 90^\circ$ for the intense-beam transport channels considered there.

In the present paper we report the results of systematic studies with different initial particle distributions at values of $\sigma_0 = 90^\circ$ and $\sigma_0 = 60^\circ$.
Of particular interest in the $\sigma_0 = 90^\circ$ case was the third-order instability.
In contrast to our results on the envelope modes, we found a much weaker response of the Gaussian distribution to the third-order instability in the cases investigated.
At the same time we discovered that the non-\mbox{K-V} distributions considered experience in the first period of  the transport channel a rapid emittance growth which occurs even in regions where no instabilities are predicted and whose growth factor increases approximately in proportion to $\sigma^{-1}$ in the strong-space-charge limit at fixed $\sigma_0$.
It will be shown that the emittance growth can be attributed to an internal redistribution of the particles toward a more uniform density profile, whereby field energy is converted into transverse kinetic energy.

\section{PROPERTIES OF DIFFERENT PARTICLE DISTRIBUTIONS\\ FOR CONTINUOUS BEAMS\label{sec2}}
\subsection{General Formulae}
Different particle distributions can have the same first and second moments.
To allow a comparison of the results obtained with different distribution functions carrying the same current, we use the concept of equivalent beams introduced by Lapostolle\cite{ref6} and Sacherer,\cite{ref7} i.e., beams having the same first and second moments.
Equality of the rms emittances alone is not sufficient for this equivalence.
Since we are dealing with continuous beams rather than bunched beams, we describe the distribution in the $4$-dimensional transverse phase space.
For the sake of simplicity, the following analytic integrals refer to a centered, round reference beam with upright, uncoupled phase-space ellipses.
We rescale the slopes according to $\xi=\ell x^\prime$ and $\eta=\ell y^\prime$, where $\ell$ is a length chosen so that all four coordinate variances are equal.
The radial phase-space density then depends only on
\begin{displaymath}
 f(r={\rm const.})={\rm const.},\qquad r^2=x^2+y^2+\xi^2+\eta^2.
\end{displaymath}
Thus $r$ is a phase-space radius, whereas $\rho=\sqrt{x^2+y^2}$ below is a real-space radius.
The scaling avoids adding unscaled positions and slopes, which have different dimensions.
Such reference distributions can be transformed linearly to the rms-matched elliptical distributions used for the quadrupole-channel simulations.
The circular electrostatic model used below is not an exact field calculation for those elliptical beams.

For five different types of particle distributions, we have evaluated
\begin{itemize}
 \item the ratio of the marginal emittance to the rms emittance,
 \item the normalized particle density $g(x,y)$ in real space,
 \item the electric field $E(\rho)$ inside the beam,
 \item the electric field energy per unit length $U$ associated with the charge distribution.
\end{itemize}

A given distribution function $f(r)$ is called normalized if
\begin{displaymath}
 \int_{r\le a} f(r)\,dV=1,
\end{displaymath}
where $dV=r^{n-1}\,dr\,d\Omega_{n-1}$ denotes the volume element of the $n$-dimensional phase space in spherical coordinates, and $a$ is the maximum value of $r$.
In our case $n=4$, $dV=dx\,dy\,d\xi\,d\eta$, and the total solid angle is $2\pi^2$.
For an untruncated Gaussian the upper radial limit is infinity.
If we assume the beam to be centered, the first moments are zero and the variance of any one normalized coordinate is given by\cite{ref8}
\begin{displaymath}
 \<x^2\>=\frac{\<r^2\>}{n}=\frac{1}{n}\int_{r\le a}r^2 f(r)\,dV\bigg/\int_{r\le a}f(r)\,dV.
\end{displaymath}
Since the particle density only depends on $r$, the integrals can be simplified to
\begin{displaymath}
 \<x^2\>=\frac{1}{n}\int_0^a f(r)\,r^{n+1}\,dr\bigg/\int_0^a f(r)\,r^{n-1}\,dr.
\end{displaymath}
The rms emittance of a distribution in the physical $x,x^\prime$-plane is defined here as\cite{ref6}
\begin{displaymath}
 \varepsilon_{\rm rms} =4\left[\<x^2\>\<{x^\prime}^2\>-\<xx^\prime\>^2\right]^{1/2}.
\end{displaymath}
We retain the notation $\<\varepsilon\>=\varepsilon_{\rm rms}$ throughout the paper.
It denotes four times the unnormalized one-rms emittance, not an average of a single-particle emittance.
If the emittance ellipse is upright, one has $\<xx^\prime\>=0$ and therefore
\begin{displaymath}
 \varepsilon_{\rm rms}=\<\varepsilon\> =4\left[\<x^2\>\<{x^\prime}^2\>\right]^{1/2}.
\end{displaymath}
The factor $4$ is chosen so that the rms emittance of a K-V distribution is equal to the area $A$ divided by $\pi$ of the minimum circumscribing phase-space ellipse that encloses all particles.
This number $A/\pi$ is called the maximum or marginal emittance $\varepsilon_{\rm max}$ of a distribution with bounded support.
For the reference coordinates, $\varepsilon_{\rm max}=a^2/\ell$ and $\varepsilon_{\rm rms}=4\<x^2\>/\ell$.
Their ratio is therefore
\begin{displaymath}
 \frac{\varepsilon_{\rm max}}{\varepsilon_{\rm rms}}=\frac{a^2}{4\<x^2\>}
 =\frac{na^2}{4}\int_0^a f(r)\,r^{n-1}\,dr\bigg/\int_0^a f(r)\,r^{n+1}\,dr.
\end{displaymath}
The density in real space $g(x,y)$ is obtained by integration over $\xi$ and $\eta$.
With $w^2=\xi^2+\eta^2$ and $\rho^2=x^2+y^2$, one obtains
\begin{displaymath}
 g(\rho)=2\pi\int_0^{\sqrt{a^2-\rho^2}}f\!\left(\sqrt{\rho^2+w^2}\right)\,w\,dw,\qquad 0\le\rho\le a.
\end{displaymath}
Here $g(\rho)=0$ for $\rho>a$, except in the untruncated Gaussian case, for which the integral over $w$ extends to infinity.
In particular, the argument of $f$ is a radius, not its square.

For the electrostatic calculation we assume a centered, circular beam in a coaxial conducting pipe of radius $R$.
Writing $\lambda=I/(c\beta)$ for the line charge, the physical charge density is $\lambda g(\rho)$, with $2\pi\int_0^a g(\rho)\rho\,d\rho=1$.
The electric field $E_i(\rho)=-d\Phi/d\rho$ inside a positively charged beam follows from Poisson's equation
\begin{eqnarray*}
 \frac{1}{\rho}\frac{d}{d\rho} \left(\rho\frac{d\Phi}{d\rho}\right)
 &=&-\frac{\lambda}{\epsilon_0}g(\rho),\\
 E_i(\rho)&=&\frac{\lambda}{\epsilon_0\rho}\int_0^\rho g(t)\,t\,dt.
\end{eqnarray*}
For a beam of finite radius $a<R$, the field between beam and wall follows from Laplace's equation.
Continuity at $\rho=a$ gives
\begin{displaymath}
 E_o(\rho)=\frac{\lambda}{2\pi\epsilon_0\rho},\qquad a\le\rho\le R.
\end{displaymath}
The electric field energy per unit longitudinal length is then
\begin{displaymath}
 U=\pi\epsilon_0\left(\int_0^a E_i^2\,\rho\,d\rho+\int_a^R E_o^2\,\rho\,d\rho\right).
\end{displaymath}
For an untruncated Gaussian there is no sharp beam boundary, and the Gaussian field must instead be integrated directly from $0$ to $R$.
Its tabulated energy is the large-$R$ limit, with an exponentially small omitted tail.

Table~\ref{table1} gives the definitions of the five particle distribution functions used in our work and the corresponding quantities.
The same rms size $s=\sqrt{\<x^2\>}$, rather than the same cutoff $a$, is to be used when comparing their field energies.
\begin{table}[p]
\centering
\refstepcounter{table}\label{table1}
\includegraphics[trim=114bp 117bp 112bp 8bp,clip,
  width=\linewidth,height=.98\textheight,keepaspectratio]{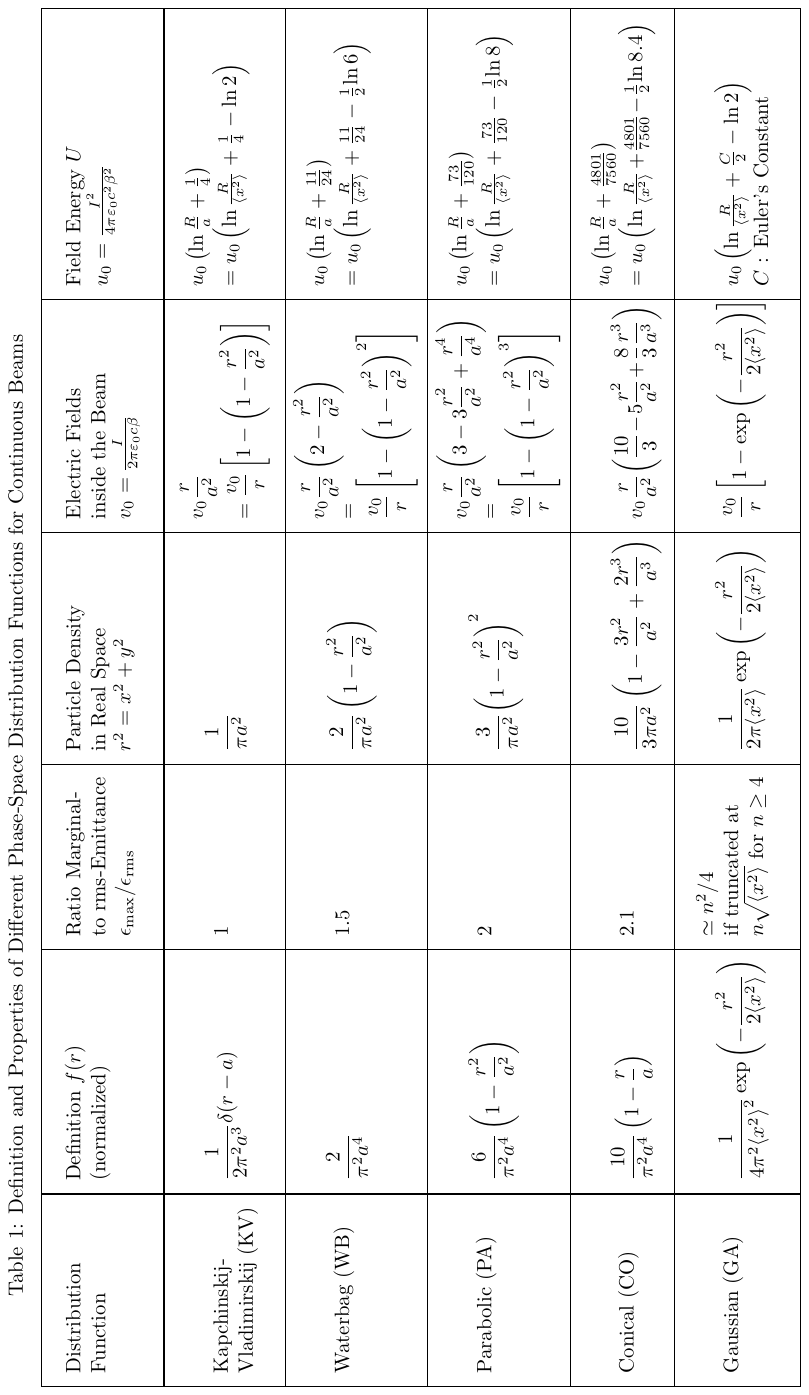}
\end{table}
\noindent
Table~\ref{table1} reproduces the supplied original artwork.
In that table, the phase-space entries denoted by $x^\prime$ and $y^\prime$ stand for the scaled coordinates $\xi$ and $\eta$ used above, and the radius in the spatial-density and electric-field columns is the real-space radius $\rho$.
The logarithms printed as $\ln(R/\<x^2\>)$ in the field-energy column are to be read as $\ln(R/\sqrt{\<x^2\>})=\ln(R/s)$, so that their arguments are dimensionless.
Here $v_0=I/(2\pi\epsilon_0c\beta)$ and $u_0=I^2/(4\pi\epsilon_0c^2\beta^2)$.
The first four distributions vanish outside their stated support.
For them $a^2/s^2=4,6,8,8.4$, respectively.
The Gaussian density, field, and field-energy entries refer to the untruncated distribution, whose marginal emittance is infinite.
Its energy expression assumes $R\gg s$, and $C$ denotes Euler's constant.
Only the approximate marginal-to-rms ratio $n_c^2/4$ refers to a finite cutoff $a=n_c s_G$, with $n_c\ge4$.
The exact normalization and second moment for that cutoff are given below.

A Gaussian truncated at a finite phase-space radius must be renormalized, and its rms moments must be recalculated before constructing an equivalent beam.
If $s_G$ is the width of the parent Gaussian, $a=n_c s_G$, and $A_c=n_c^2/2$, the retained probability and the variance after renormalization are
\begin{displaymath}
 D=1-(1+A_c)e^{-A_c},\qquad
 \<x^2\>_{\rm tr}=s_G^2\frac{1-(1+A_c+A_c^2/2)e^{-A_c}}{D}.
\end{displaymath}
The truncated phase-space density is the parent Gaussian divided by $D$ for $r\le a$ and zero outside.
Its exact marginal-to-rms ratio is $a^2/(4\<x^2\>_{\rm tr})$.
For $n_c=4$, $D\simeq0.996981$ and $\<x^2\>_{\rm tr}\simeq0.989233s_G^2$.
Thus the Gaussian entries in Tables~\ref{table1} and~\ref{table2} are analytic untruncated reference values, not exact finite-cutoff values.

\clearpage
\subsection{Differences of the Field Energy Associated with Different Distribution Functions}
As can be seen in Table~\ref{table1}, each charge distribution function is associated with a specific electric field energy.
At fixed line charge, fixed transverse second moments, and the same enclosing pipe, the homogeneous circular density is the minimum-energy reference for the circular profiles considered here.
It is the spatial density associated with a K-V distribution, although a homogeneous spatial density does not uniquely determine the full phase-space distribution.

Assuming that each distribution has the same spatial second moments, as stated earlier, one finds that the field energy $U$ increases from the K-V to the Gaussian distribution.
One can write the difference between a particular distribution and the K-V distribution in the form
\begin{displaymath}
 \Delta U=U-U_{\rm KV}=f \cdot u_0,
\end{displaymath}
where
\begin{displaymath}
 u_0=\frac{I^2}{4\pi\epsilon_0c^2\beta^2}
 \simeq10^{-7}(I/\beta)^2\quad[\,{\rm joules/meter}\,],
\end{displaymath}
where the numerical expression uses $I$ in amperes.
The scalar $f$ in $\Delta U=f u_0$ is an energy-difference factor, distinct from the radial density function $f(r)$.
The factors $f$ for the field-energy difference between each distribution and the equivalent K-V beam are listed in Table~\ref{table2}.
As an example, one finds for the additional field energy stored in a Gaussian distribution compared to an equivalent K-V beam
\begin{displaymath}
 \Delta U=U_{\rm GA}- U_{\rm KV}=u_0\cdot(\half C-\oqu)=u_0f,
\end{displaymath}
with
\begin{displaymath}
 f=\half C-\oqu\simeq 0.0386,
\end{displaymath}
wherein $C\simeq 0.5772$ denotes Euler's constant.
\begin{table}[ht]
  \begin{center}
    \caption{$f$-Factors for Field-Energy Differences between Equivalent Charge Distributions}\medskip
    \begin{tabular}{c|cccc}\hline
      \hphantom{from}to & KV & WB & PA & CO\\
      from &&&&\\ \hline
      ~WB  & 0.00560 & 0 & --- & ---\\
      ~PA  & 0.01176 & 0.00616 & 0 & ---\\
      ~CO  & 0.01408 & 0.00848 & 0.00232 & 0 \\
      ~GA  & 0.03861 & 0.03301 & 0.02685 & 0.02452\\ \hline
    \end{tabular}
    \label{table2}
  \end{center}
\end{table}

As will be explained in the next section, we surmise that this additional field energy is converted into particle kinetic and potential energy (and hence emittance growth) as the distribution tends to become more homogeneous.
For particles of positive charge $\zeta e_0$, with $e_0$ the elementary charge, the electrostatic energy difference per particle is
\begin{displaymath}
 \Delta E_{\rm el}=\frac{fu_0}{N}=\frac{fI\zeta e_0}{4\pi\epsilon_0c\beta},\qquad N=\frac{I}{\zeta e_0\beta c}.
\end{displaymath}
Here $N$ is the number of particles per unit longitudinal length and $\Delta E_{\rm el}$ is an energy in joules.
Expressed numerically in electronvolts, it is approximately $30fI\zeta/\beta$ when $I$ is in amperes.
The effective energy change relevant to transverse motion also includes the self-magnetic reduction by $\gamma^{-2}$, as used below.
The electric field energy $U$ itself does not include the magnetic field energy.

\subsection{Emittance Growth Due to a Change of the Particle Distribution\label{sec:2.3}}
\begin{figure}[tp]
\centering\epsfig{file=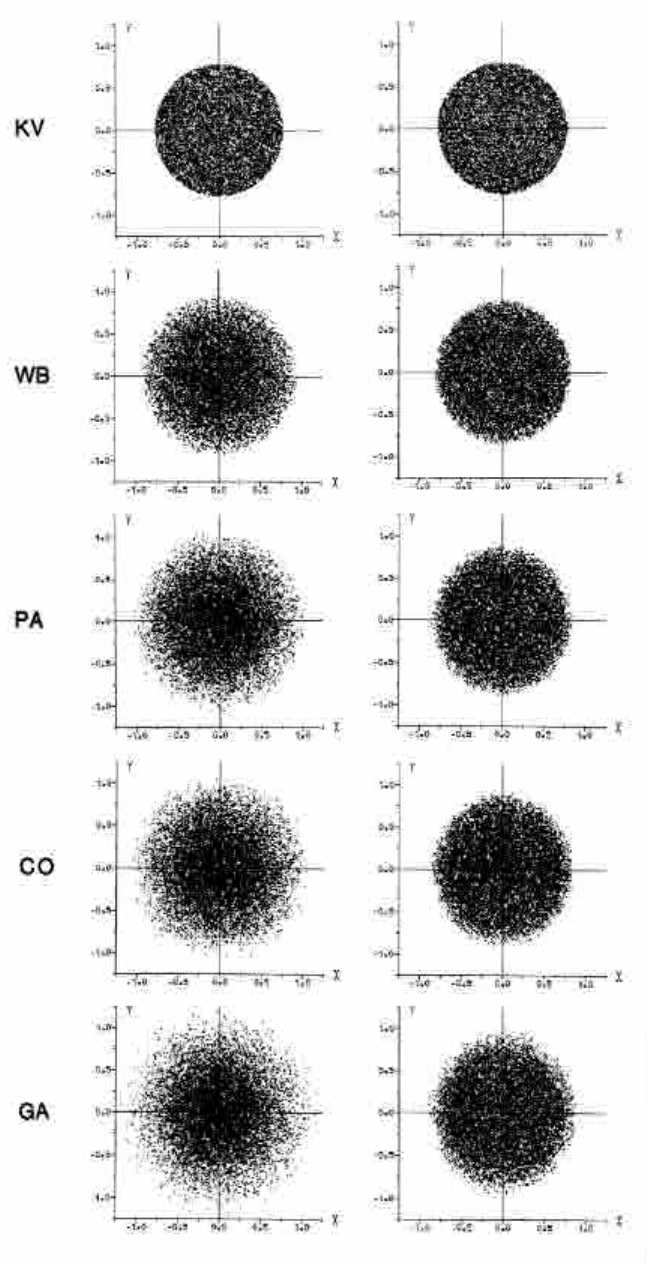,width=.6\linewidth,angle=0}
\caption{Real-space $(x,y)$ projections of K-V, waterbag, parabolic, conical and Gaussian distributions before and after
the first cell of the GSI FODO channel at $ \sigma_0=60^\circ$, $\sigma=15^\circ$.}
\label{fig1}
\end{figure}

The non-K-V distributions constructed above are not, in general, matched self-consistent distributions of the periodic focusing channel.
They undergo an initial redistribution toward a more homogeneous charge density in real space, as illustrated in Fig.~\ref{fig1} after one period of the GSI quadrupole channel, whose parameters are defined in Ref.~5.
This rearrangement can convert excess self-field energy into transverse kinetic energy and energy associated with the beam's size in the focusing field.
A more uniform spatial density does not imply that the final phase-space distribution is a K-V distribution or an exactly stationary state.
Moreover, increases of spatial or angular second moments alone do not prove emittance growth, because $\<xx^\prime\>$ also enters its definition.
The observed growth concerns the projected rms emittance and does not imply an increase of the fine-grained phase-space volume in collisionless Hamiltonian motion.

The following estimates use a round-beam smooth approximation to the periodic channel, neglect particle loss and acceleration, and assume equal sharing between the two transverse planes.
Complete removal of the initial excess energy $f u_0$ is an idealization.
Incomplete redistribution, changes in rms size, and coherent correlations can modify the result.
The two prescriptions are intended as limiting energy-partition models, not as exact conservation laws for the self-consistent beam.

\bigskip
\noindent
$(a)\,\,\, \sigma\to\sigma_0$\quad Harmonic, rematched-beam estimate

\bigskip
\noindent
In order to obtain an analytical estimate of the emittance growth for low intensities, an rms-matched non-K-V distribution that changes its charge density in real space towards a homogeneous one is treated as an equivalent K-V distribution  that becomes ``heated up''.
The effective heating is assumed to increase both the mean kinetic energy, proportional to $\<{x^\prime}^2\>$, and the mean potential energy, proportional to $\<x^2\>$, with a rematched equivalent beam after redistribution.
The mean single-particle oscillator energy $E_{\rm tot}$ in one transverse plane can be expressed in terms of the depressed phase advance $\sigma$, the length of one focusing period $S$, and the transverse rms emittance $\<\varepsilon\>$ as
\begin{displaymath}
 E_{\rm tot}=\oqu\gamma m_0c^2\beta^2\sigma \<\varepsilon\>/S.
\end{displaymath}
Here the longitudinal motion may be relativistic, with Lorentz factor $\gamma$, while the transverse motion is paraxial and its kinetic energy is expanded to quadratic order.
The transverse inertial mass is then $\gamma m_0$.
The mean effective energy gain per particle in one transverse plane is modeled as
\begin{displaymath}
 \Delta E_x=\frac{\Delta E_{\rm el}}{2\gamma^2}=\frac{fI\zeta e_0}{8\pi\epsilon_0c\beta\gamma^2}.
\end{displaymath}
The factor $\half$ expresses the assumed equal partition between the two transverse planes.
The factor $\gamma^{-2}$ accounts for the cancellation of part of the transverse electric self-force by the magnetic self-force of a beam with a common longitudinal velocity.
If the superscript $^+$ denotes the appropriate quantities after the homogenization of the charge distribution, we obtain for the total energies
\begin{displaymath}
 \frac{\gamma m_0c^2\beta^2}{4S}\sigma\<\varepsilon\>+\frac{fI\zeta e_0}{8\pi\epsilon_0c\beta\gamma^2}
 =\frac{\gamma m_0c^2\beta^2}{4S}\sigma^+\<\varepsilon\>^+.
\end{displaymath}
Thus
\begin{displaymath}
 \sigma\<\varepsilon\>+\frac{2fIS}{I_0\beta^3\gamma^3}=\sigma^+\<\varepsilon\>^+,
\end{displaymath}
where $I_0=4\pi\epsilon_0m_0c^3/(\zeta e_0)$ is the characteristic current for the chosen particle species.
For singly charged particles it reduces to the definition used in Ref.~5.
Since $\sigma$ does not remain constant during reorganization of the particle distribution but increases to $\sigma^+$, we have to express $\sigma^+$ in terms of $\<\varepsilon\>^+$.
According to the smooth-approximation theory of Ref.~9, this means
\begin{displaymath}
 \sigma^+=\sigma_0\left(\sqrt{1+{u^+}^2}-u^+\right),\qquad
 u^+=\frac{IS}{I_0\beta^3\gamma^3\sigma_0\<\varepsilon\>^+}.
\end{displaymath}

Solving for the ratio $\<\varepsilon\>^+/\<\varepsilon\>$, one obtains after some algebraic transformations
\begin{displaymath}
 \frac{\<\varepsilon\>^+}{\<\varepsilon\>}=
\sqrt{\left(\frac{\sigma}{\sigma_0}+2uf\right)\left(\frac{\sigma}{\sigma_0}+2uf+2u\right)}
\end{displaymath}
with $u=IS/(I_0\beta^3\gamma^3\sigma_0\<\varepsilon\>)$ pertaining to the initial rms emittance $\<\varepsilon\>$.
For example, writing $\chi=\<\varepsilon\>^+/\<\varepsilon\>$, the energy prescription gives $\sigma^+\chi/\sigma_0=\sigma/\sigma_0+2uf$,
while the matched-beam relation gives $(\sigma^+\chi/\sigma_0)^2+2u\sigma^+\chi/\sigma_0=\chi^2$.
$u$ can easily be expressed in terms of $\sigma_0$ and $\sigma$ as
\begin{displaymath}
 2u=\frac{\sigma_0}{\sigma}-\frac{\sigma}{\sigma_0}.
\end{displaymath}
The formula for the relative growth of the transverse rms emittance then reduces to
\begin{equation}\label{eq:formula1}
 \frac{\<\varepsilon\>^+}{\<\varepsilon\>}\simeq
 \sqrt{\left[1+f\left(\frac{\sigma_0^2}{\sigma^2}-1\right)\right]
       \left[1+f\left(1-\frac{\sigma^2}{\sigma_0^2}\right)\right]}.
\end{equation}
The algebraic result is exact within this prescription, but its application to the actual beam is approximate.
In particular, the mean oscillator energy is not the conserved total beam energy.
If the rms size changes, the reference K-V field energy changes as well.
Equation~\eqref{eq:formula1} does not constitute an exact energy-conservation treatment of that change.

\bigskip
\noindent
$(b)\,\,\, \sigma\to0$\quad Fixed-size, kinetic-energy estimate

\bigskip
\noindent
For non-K-V distributions the effective self-consistent potential is generally not harmonic.
In a strongly space-charge-dominated beam that has become nearly uniform in its central region, the external focusing and the self-force approximately cancel there.
A steep confining edge then motivates a ``reflecting-wall potential'' model, in which particles move almost force-free in the interior and are reflected near the boundary.
This is an idealization of the redistributed beam, not a general description of the initial peaked profiles.
For a second estimate, all the available effective energy is assigned to transverse kinetic energy while the rms size is held fixed.
The harmonic equivalent-beam relation is used only to express the initial moments in terms of $\sigma$ and $\<\varepsilon\>$.
Since the mean transverse kinetic energy of one particle is given by
\begin{displaymath}
 E_k=\half\gamma m_0c^2\beta^2\< {x^\prime}^2\>,
\end{displaymath}
the energy equation may then be written as
\begin{eqnarray*}
 \half\gamma m_0c^2\beta^2\<{x^\prime}^2\>+\frac{fI\zeta e_0}{8\pi\epsilon_0c\beta\gamma^2}
 &=&\half\gamma m_0c^2\beta^2\<{x^\prime}^2\>^+,\\
 \<{x^\prime}^2\>+\frac{fI}{I_0\beta^3\gamma^3}
 &=&\<{x^\prime}^2\>^+.
\end{eqnarray*}
If $\<x^2\>^+=\<x^2\>$ and the ellipse is upright both before and after the idealized energy transfer, so that $\<xx^\prime\>=\<xx^\prime\>^+=0$, we obtain
\begin{displaymath}
 \big(\<\varepsilon\>^+\big)^2=\<\varepsilon\>^2+\frac{16fI\<x^2\>}{I_0\beta^3\gamma^3}.
\end{displaymath}
With the relation
\begin{displaymath}
 \< x^2\> =\oqu\< \varepsilon\> S/\sigma,
\end{displaymath}
which is the matched smooth-approximation relation for the initial equivalent K-V beam, this finally leads to
\begin{equation}\label{eq:formula2}
 \frac{\<\varepsilon\>^+}{\<\varepsilon\>}\simeq\sqrt{1+2f\left(\frac{\sigma_0^2}{\sigma^2}-1\right)}.
\end{equation}
For the values of $f$ considered here and $0<\sigma\le\sigma_0$, the estimate in Eq.~\eqref{eq:formula1} does not exceed that in Eq.~\eqref{eq:formula2}.
This ordering of two model expressions is not a proof that the actual growth lies between them.
Both reduce to unity for $f=0$ or $\sigma=\sigma_0$.
At fixed nonzero $f$, their strong-space-charge growth factors are proportional to $\sigma_0/\sigma$, and the case $\sigma=0$ itself is a singular limit.

\section{SIMULATION STUDIES\label{sec3}}
\subsection{Third-Order Instability with Equivalent Beams of Different Particle Distributions}
The space-charge potential of an ideal K-V distribution is a quadratic function of the spatial coordinates inside the beam, leading to linear self-forces.
As has been shown previously, the K-V distribution is unstable against perturbations of this potential distribution in some specific regions depending on the external forces (defined by the phase advance $\sigma_0$) and the beam parameters (defined by $\sigma$).
The type of perturbation can be classified by the order of additional potential terms.
For example, a potential proportional to $xy^2$ is called a ``third-order'' perturbation potential.
For $\sigma_0=90^\circ$, the K-V distribution is unstable against this type of perturbation in the region $38^\circ\le \sigma\le 60^\circ$.
To evaluate whether and how this instability affects the non-K-V distributions defined above, we performed simulation studies for $\sigma_0=90^\circ$ and $\sigma =41^\circ$, hence near the maximum growth rate for the third-order mode in a K-V beam.
All computer runs discussed in this paper were made with the parameters of the GSI quadrupole channel (see Ref.~5).
The conclusions below refer to the reported runs and propagation intervals.
A numerical convergence study is not documented here.
Independent verification of small growth rates would require the particle number, space-charge mesh, integration steps, initialization and matching procedure, perturbation seed, and treatment of particle loss.

\begin{figure}[H]
\centering\includegraphics[angle=-90,width=\linewidth]{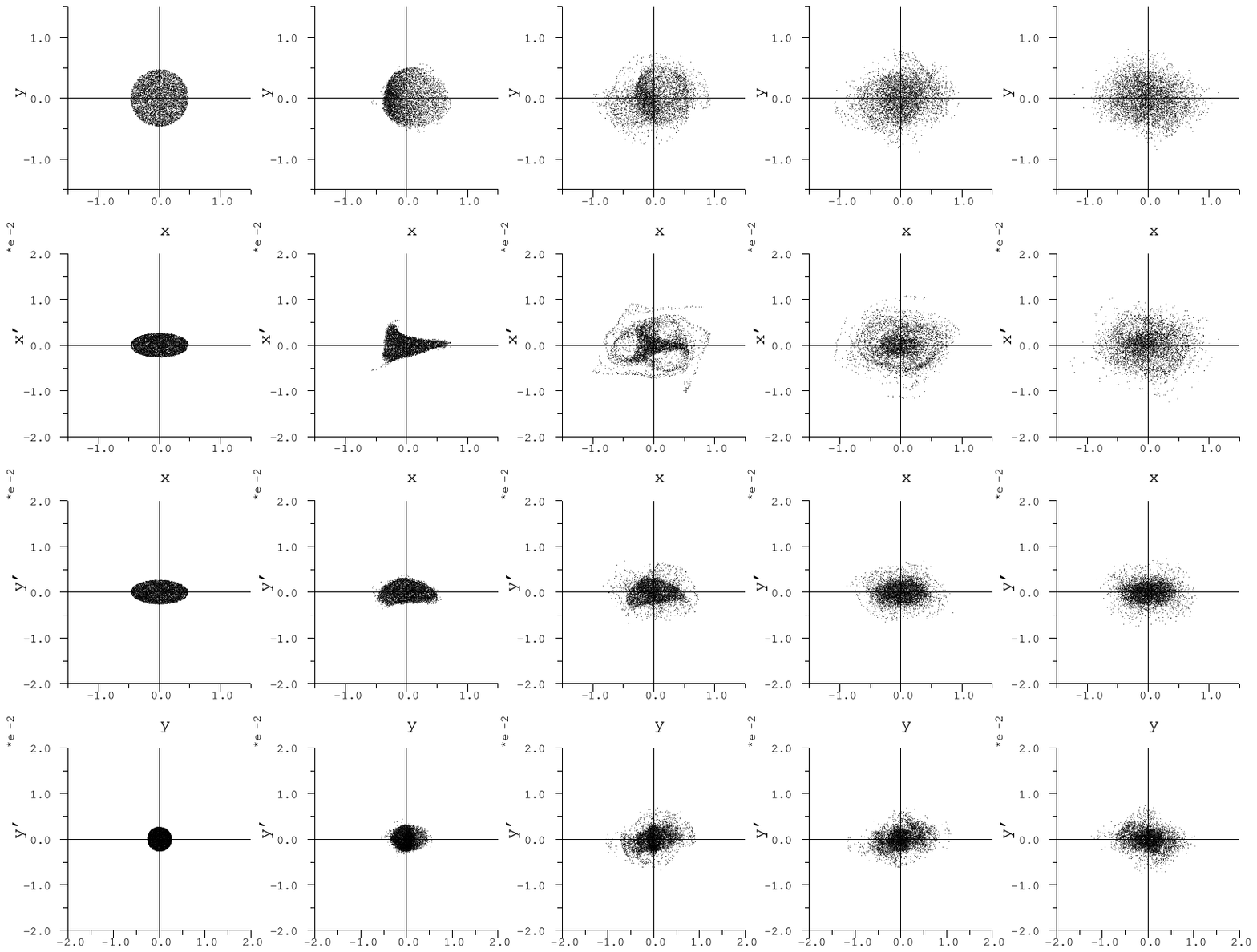}
\\~\\ Period\hspace*{5mm} \!0\,\hspace*{2cm} 20\hspace*{2cm} 40\hspace*{2cm} 60\hspace*{2cm} \!100\hspace*{\fill}
\caption{Transformation of an initial K-V distribution through the GSI FODO channel at $\sigma_0=90^\circ, \sigma=41^\circ$.
(The $x,x^\prime$ and $y,y^\prime$ phase-space projections are represented in normal form.)}
\label{fig2}
\end{figure}

Figure~\ref{fig2} shows the evolution of an initial K-V distribution.
The specific instability can easily be recognized by the three ``arms'' growing out of the initial elliptic particle distribution.
After about $50$ periods the transverse rms  emittance has grown to a factor of $2.2$, remaining nearly constant during the following periods.

\begin{figure}[H]
\centering\includegraphics[angle=-90,width=\linewidth]{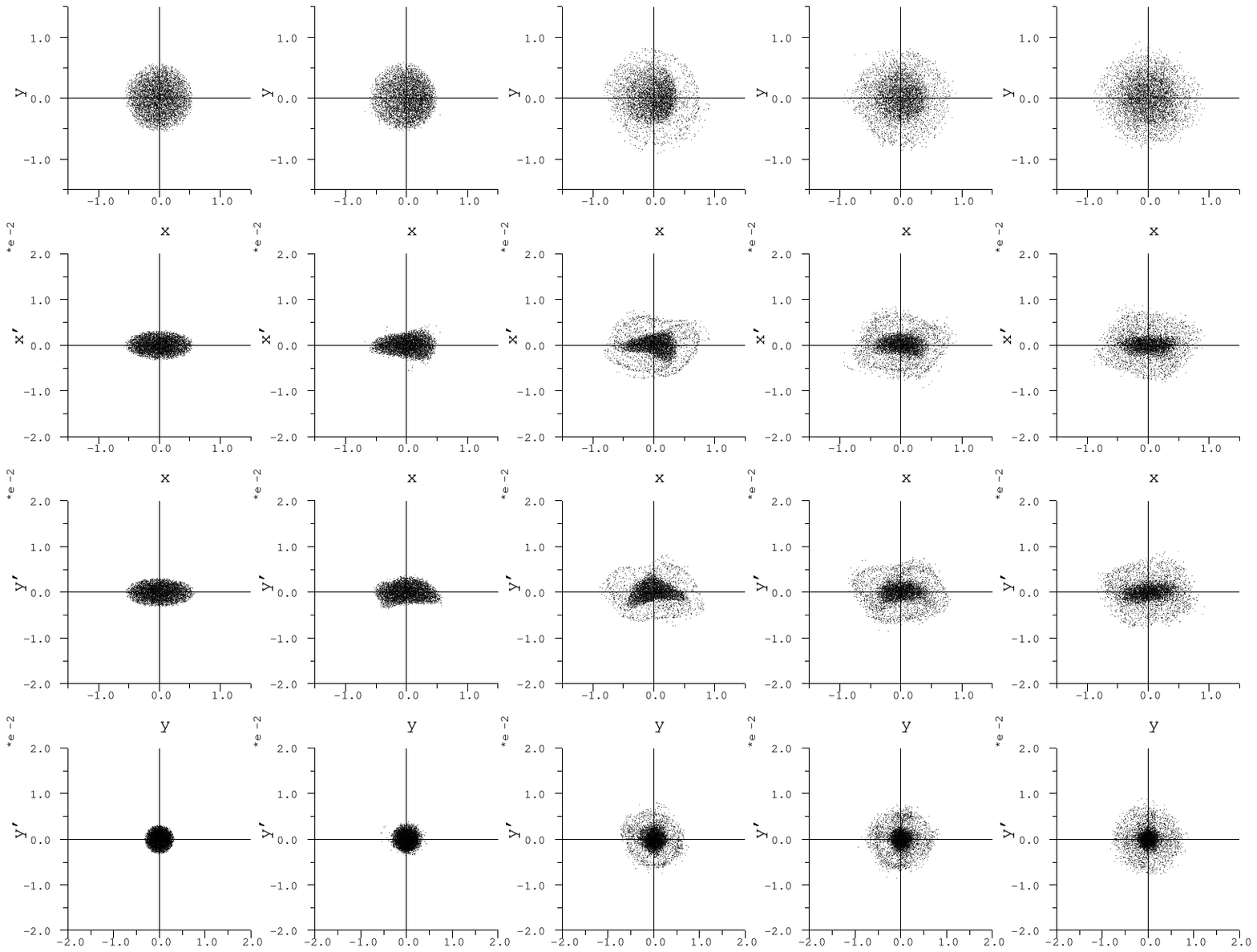}
\\~\\ Period\hspace*{5mm} \!0\,\hspace*{2cm} 20\hspace*{2cm} 40\hspace*{2cm} 60\hspace*{2cm} \!100\hspace*{\fill}
\caption{Transformation of an initial waterbag distribution through the GSI FODO channel at $\sigma_0=90^\circ, \sigma=41^\circ$.
(The $x,x^\prime$ and $y,y^\prime$ phase-space projections are represented in normal form.)}
\label{fig3}
\end{figure}

Figure~\ref{fig3} shows the transformation of an equivalent initial waterbag distribution under the same conditions ($\sigma_0=90^\circ$, $\sigma =41^\circ$).
The resulting type of distortion is obviously the same, yet the growth of ``arms'' is less pronounced in that case.
The rms emittance growth factor of $2.2$ after $100$ periods shows no essential difference to that of the K-V distribution.

\begin{figure}[H]
\centering\includegraphics[angle=-90,width=\linewidth]{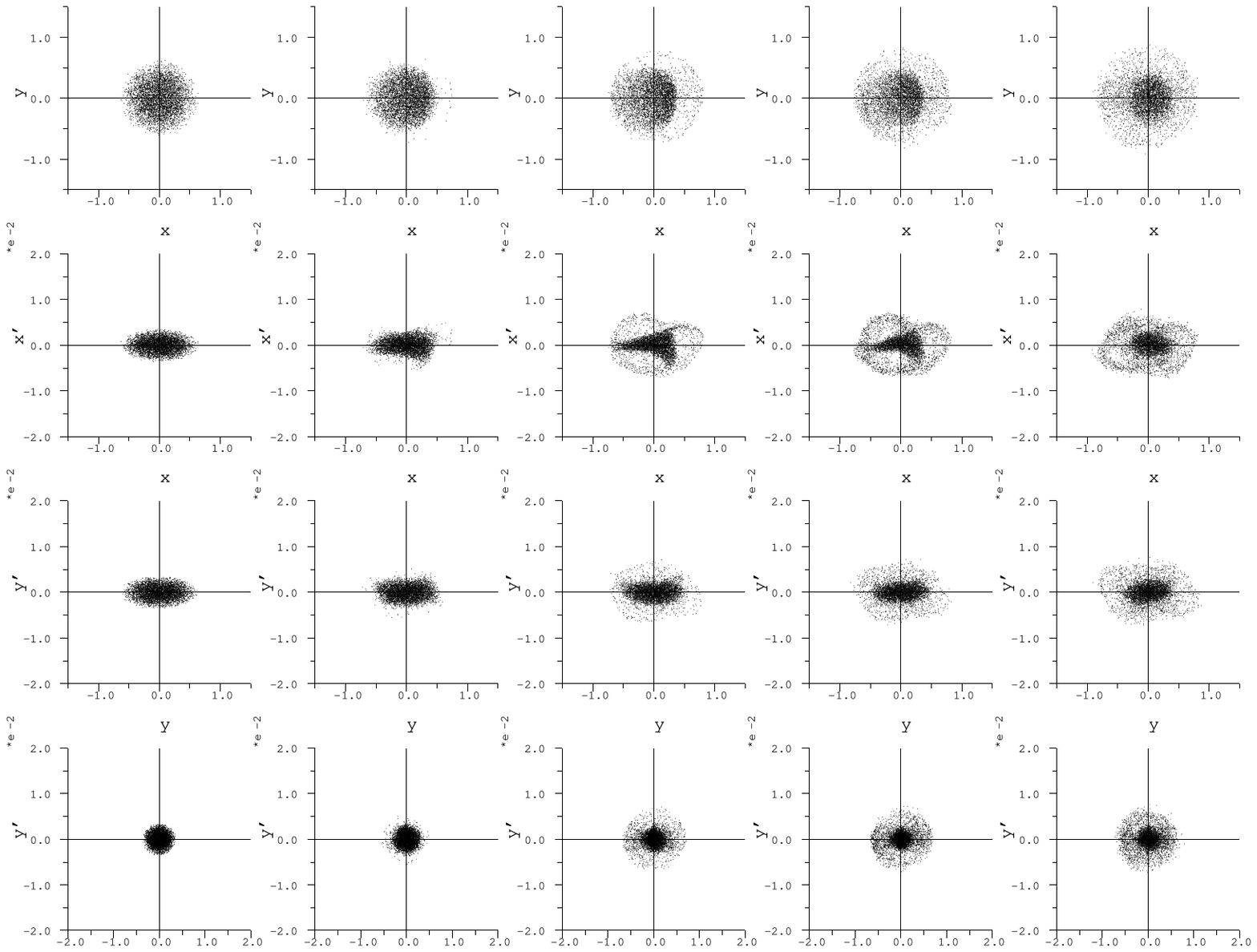}
\\~\\ Period\hspace*{5mm} \!0\,\hspace*{2cm} 20\hspace*{2cm} 40\hspace*{2cm} 60\hspace*{2cm} \!100\hspace*{\fill}
\caption{Transformation of an initial parabolic distribution through the GSI FODO channel at $\sigma_0=90^\circ, \sigma=41^\circ$.
(The $x,x^\prime$ and $y,y^\prime$ phase-space projections are represented in normal form.)}
\label{fig4}
\end{figure}

Figure~\ref{fig4} shows the evolution of an initial parabolic phase-space distribution, whose emittance plots display patterns similar to those of the waterbag distribution,
yielding a saturated transverse rms emittance growth of $1.8$ after $100$ periods.

\begin{figure}[H]
\centering\includegraphics[angle=-90,width=\linewidth]{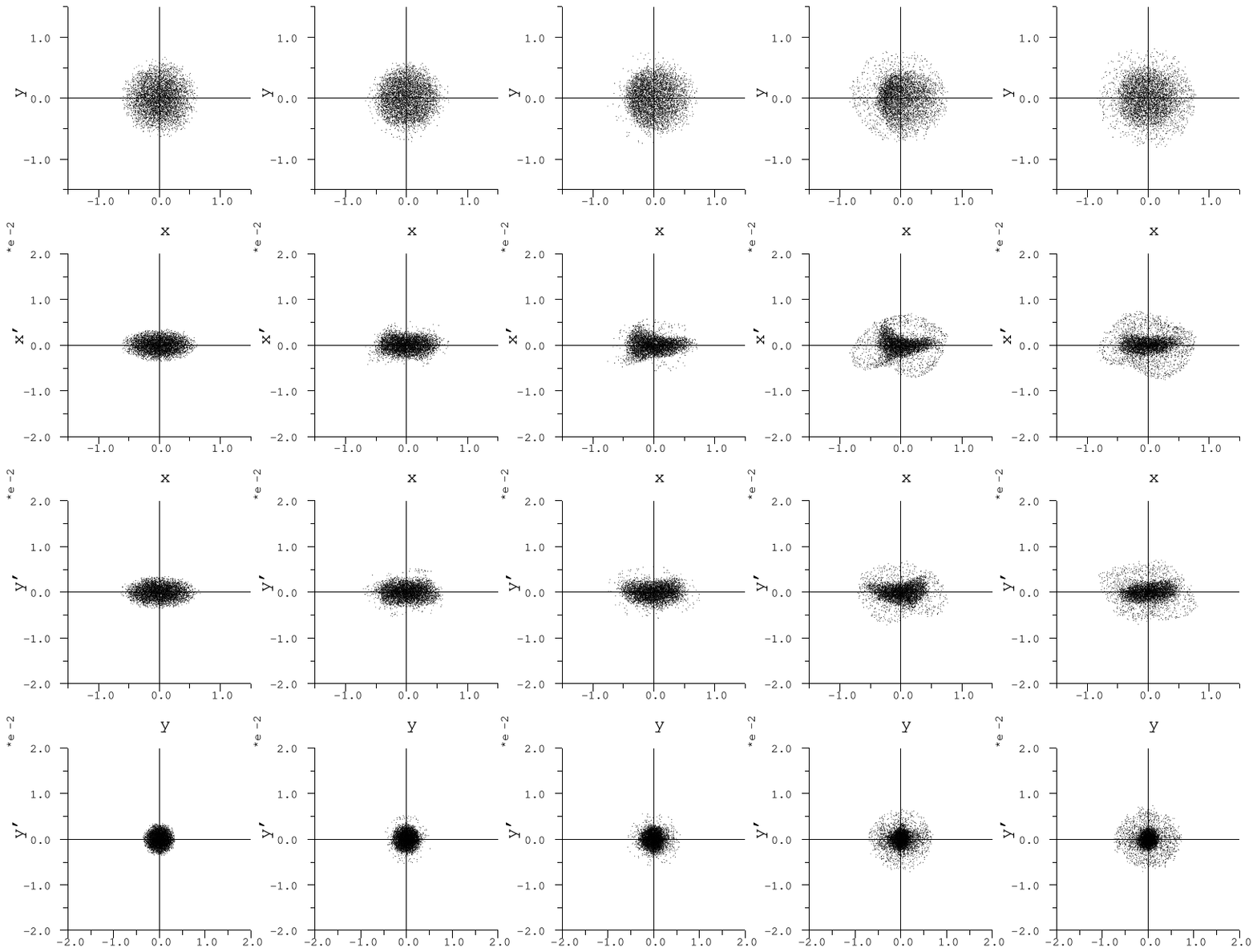}
\\~\\ Period\hspace*{5mm} \!0\,\hspace*{2cm} 20\hspace*{2cm} 40\hspace*{2cm} 60\hspace*{2cm} \!100\hspace*{\fill}
\caption{Transformation of an initial conical distribution through the GSI FODO channel at $\sigma_0=90^\circ, \sigma=41^\circ$.
(The $x,x^\prime$ and $y,y^\prime$ phase-space projections are represented in normal form.)}
\label{fig5}
\end{figure}

The evolution of the conical distribution, plotted in Fig.~\ref{fig5}, shows a less pronounced third-order instability mode compared to the previous types of distributions.
The growth factor of the rms emittance is lower than in these cases reaching a value of $\simeq 1.75$ after $100$ periods.

\begin{figure}[H]
\centering\includegraphics[angle=-90,width=\linewidth]{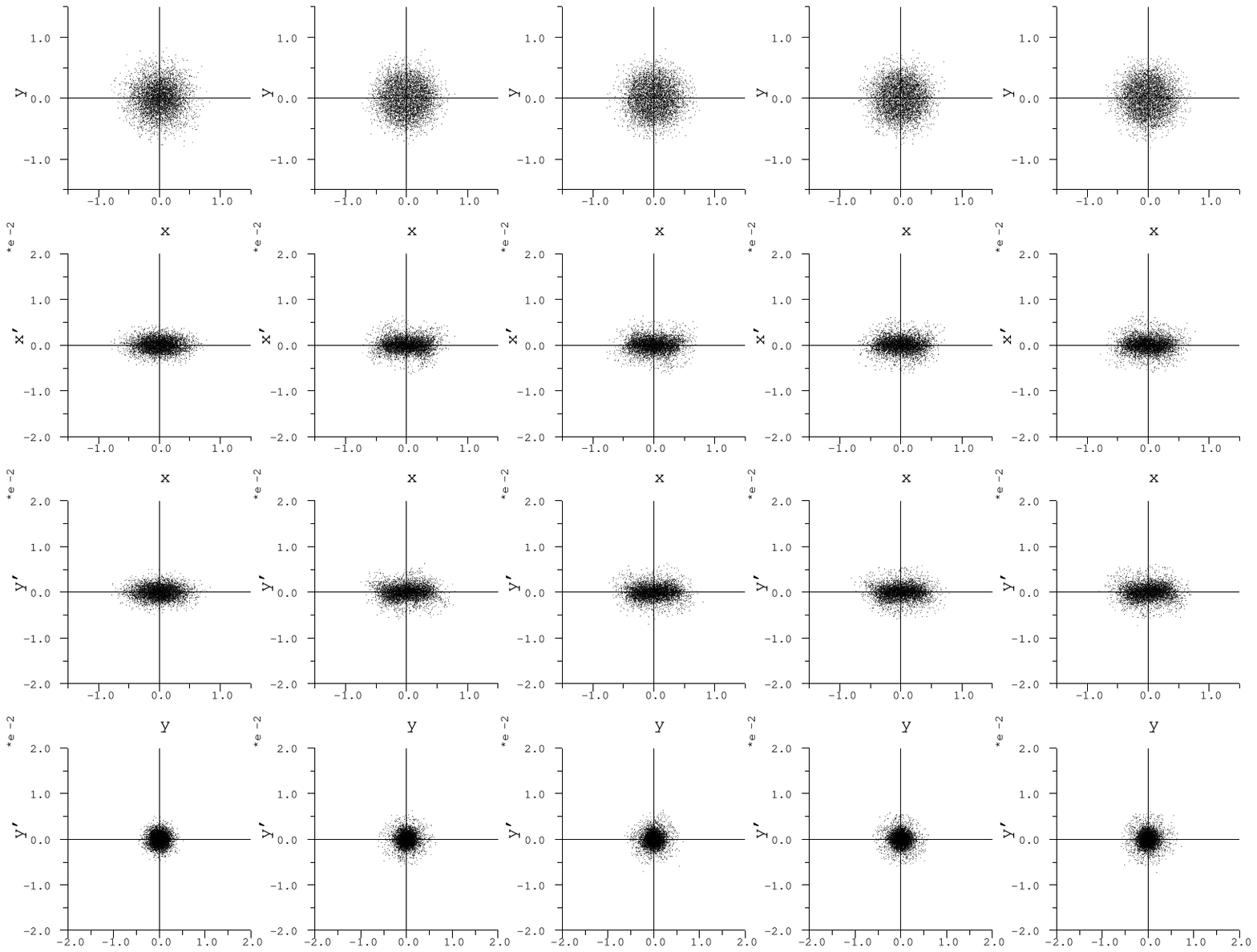}
\\~\\ Period\hspace*{5mm} \!0\,\hspace*{2cm} 20\hspace*{2cm} 40\hspace*{2cm} 60\hspace*{2cm} \!100\hspace*{\fill}
\caption{Transformation of an initial Gaussian distribution through the GSI FODO channel at $\sigma_0=90^\circ, \sigma=41^\circ$.
(The $x,x^\prime$ and $y,y^\prime$ phase-space projections are represented in normal form.)}
\label{fig6}
\end{figure}

The last density distribution investigated here is the Gaussian distribution, whose evolution is shown in Fig.~\ref{fig6}.
Its cutoff is stated in the original simulation specification as $a=4\sqrt{\<x^2\>}$.
A finite-cutoff implementation must distinguish the parent Gaussian width from the actual rms width and apply the moment correction described in Section~2.1 when constructing an equivalent beam.
A nearly constant slight increase in rms emittance is obtained and no saturation of that growth can be recognized after $100$ periods, where the growth factor of the rms emittance amounts to only $\simeq 1.18$.

\begin{figure}[H]
\centering\epsfig{file=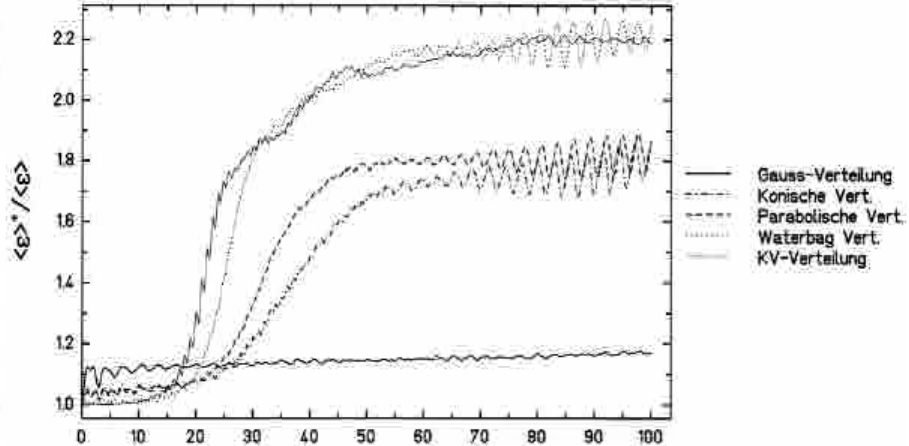,width=\linewidth,angle=0}
\\ \hspace*{-15mm}{\bf Cell Number}
\caption{Emittance growth factors versus the number of cells obtained from particle simulations for initial K-V, waterbag, parabolic, conical,
and Gaussian distributions at $\sigma_0=90^\circ$, $\sigma=41^\circ$.}
\label{fig7}
\end{figure}

Figure~\ref{fig7} shows the increase of the transverse rms emittance versus the number of cells for our five types of distribution functions.
The initial offset of the non-K-V distributions is consistent with the homogenization of the particle density in real space discussed in the previous section.
According to the emittance-growth formulae derived in Section~\ref{sec:2.3}, which give two limiting model estimates for this effect,
this leads to an initial growth of $1.3\%$ to $2.1\%$ for the waterbag, $2.7\%$ -- $4.4\%$ for the parabolic, $3.2\%$ -- $5.2\%$ for the conical,
and $8.7\%$ -- $13.8\%$ for the Gaussian distribution.
This agrees with the simulation results plotted in Fig.~\ref{fig7}, since the obtained initial emittance growth factors lie between these values.

The K-V and waterbag curves saturate near a growth factor of $2.2$, while the conical and Gaussian curves remain at lower levels over the reported interval.
The slope of the emittance growth decreases from the K-V towards the Gaussian distribution, and therefore the number of cells, where a saturation can be observed, if there is any, increases.
A possible explanation is a spread in the individual particle phase advances and the associated phase mixing, which can reduce the coherent response.
The emittance curves alone do not establish this mechanism or provide a measurement of a damping rate.
This interpretation remains to be investigated.

\begin{figure}[H]
\centering\epsfig{file=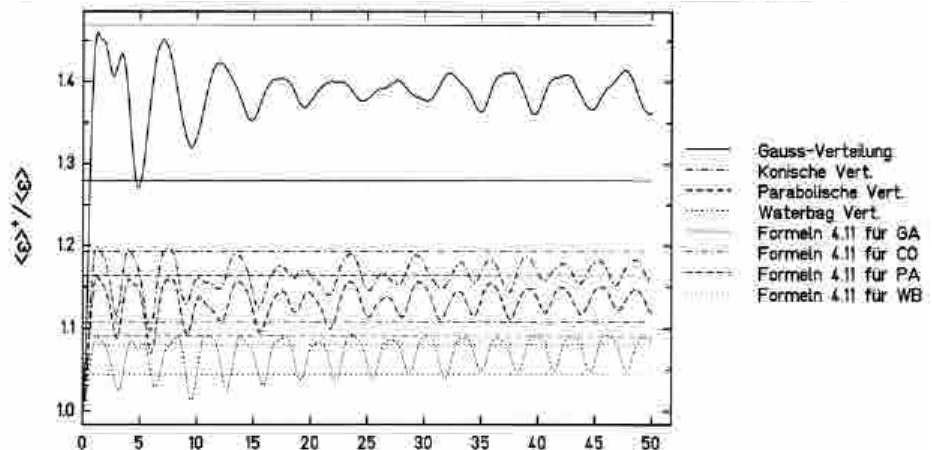,width=\linewidth,angle=0}
\\ \hspace*{-15mm}{\bf Cell Number}
\caption{Emittance growth factors versus the number of cells obtained from particle simulations for initial K-V, waterbag, parabolic, conical,
and Gaussian distributions at $\sigma_0=60^\circ$, $\sigma=15^\circ$. Upper and lower limits are obtained from Eqs.~(\ref{eq:formula1}) and~(\ref{eq:formula2}).}
\label{fig8}
\end{figure}

For comparison, the original numerical summary gives $\sigma_0=60^\circ$, $\sigma=15^\circ$ for the case associated with Fig.~\ref{fig8}.
No significant resonance-driven secular emittance growth is reported under these conditions.
The reported initial increases are approximately $1.7\%$ for the waterbag, $4.0\%$ for the parabolic, $5.0\%$ for the conical, and $12.5\%$ for the Gaussian distribution.
They lie between the corresponding model estimates, which are $1.56$--$2.63\%$, $3.26$--$5.45\%$, $3.90$--$6.49\%$, and $10.52$--$16.94\%$, respectively.

\subsection{K-V and Gaussian Distributions at Increasing Intensities\label{sec:3.2}}
Using the Gaussian distribution as a model of a smooth, centrally peaked beam, we investigated its behavior over a wide parameter range of decreasing $\sigma$ values
(i.e., increasing intensity) and compared it with a K-V distribution.
We made a large number of simulation runs for the two distributions at fixed values of $\sigma_0=90^\circ$ and $\sigma_0=60^\circ$ using as before the geometry of the GSI channel.

\bigskip\noindent
$(a)\,\,\,\sigma_0=90^\circ$

\begin{figure}[H]
\centering\epsfig{file=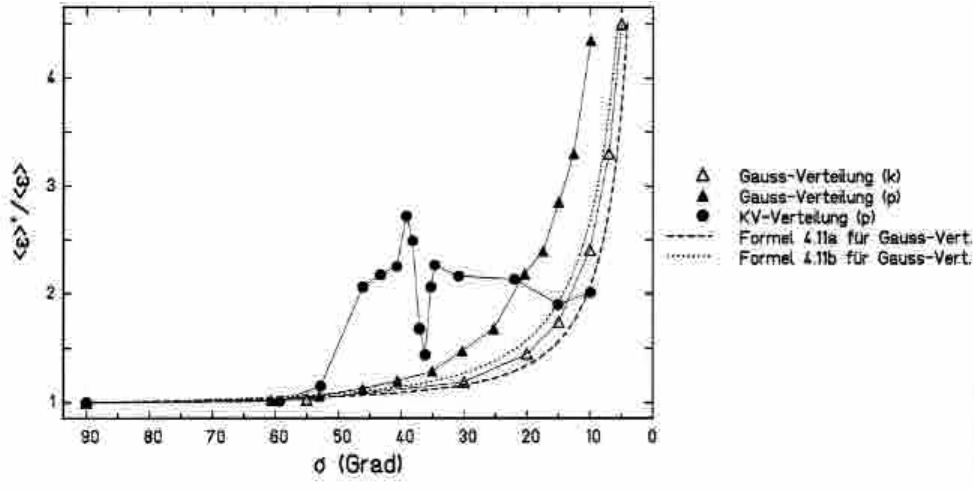,width=.79\linewidth,angle=0}
\caption{Emittance growth factors versus $\sigma$ after $100$ cells for initial K-V and Gaussian distributions at $\sigma_0=90^\circ$.
Upper and lower limits are obtained from Eq.~(\ref{eq:formula1}) (dashed curve) and~(\ref{eq:formula2}) (dotted curve).}
\label{fig9}
\end{figure}

\bigskip\noindent
Figure~\ref{fig9} shows the rms emittance growth after $100$ cells for a K-V and a Gaussian distribution at $\sigma_0=90^\circ$ and various values of $\sigma$.
As can be seen, the two types of distribution functions behave quite differently:
\begin{itemize}
 \item The K-V distribution shows the expected emittance growth due to a third-order instability in the region $60^\circ\ge\sigma\ge 38^\circ$.
 The sharp decrease of the emittance growth at $\sigma\simeq 38^\circ$ indicates that the third-order instability is no longer present.
 As $\sigma$ decreases, i.e., at higher intensities, the fourth- and higher-order modes take effect in the region $\sigma< 37^\circ$.
 \item The Gaussian distribution shows increasing emittance growth as $\sigma$ decreases.
 One part of the emittance growth can be attributed to the homogenization effect discussed in Section~\ref{sec:2.3}, since a sharp rise in rms emittance occurs within the first cell.
 The two analytical estimates are shown by the reference curves corresponding to Eqs.~\eqref{eq:formula1} and~\eqref{eq:formula2}.
 In part of the parameter range, the simulated growth exceeds that accounted for by the initial-redistribution models.
 This difference suggests that the action of the resonances in the region $\sigma<60^\circ$ is still present.
\end{itemize}

\bigskip\noindent
$(b)\,\,\,\sigma_0=60^\circ$

\begin{figure}[H]
\centering\epsfig{file=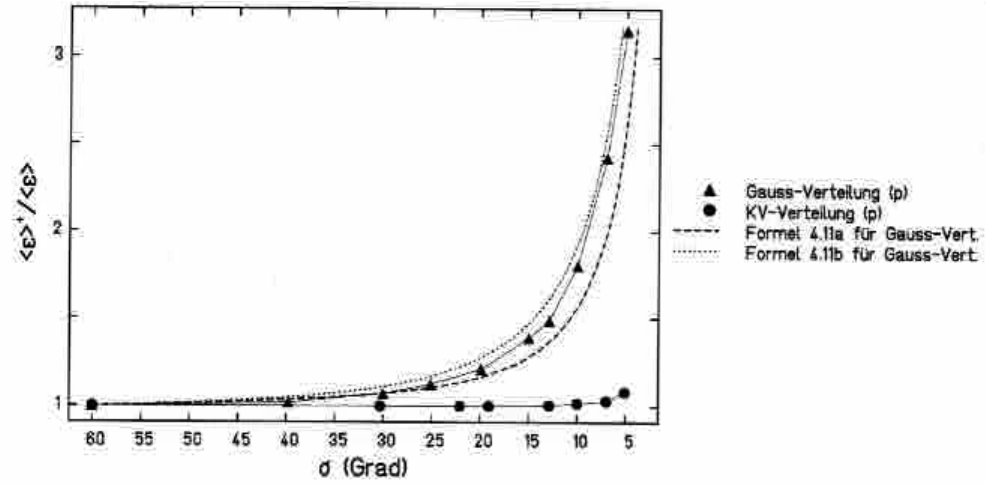,width=.79\linewidth,angle=0}
\caption{Emittance growth factors versus $\sigma$ after $50$ cells for initial K-V and Gaussian distributions at $\sigma_0=60^\circ$.
Upper and lower limits are obtained from Eq.~(\ref{eq:formula1}) (dashed curve) and~(\ref{eq:formula2}) (dotted curve).}
\label{fig10}
\end{figure}

\bigskip\noindent
Figure~\ref{fig10} shows the rms emittance growth after $50$ cells for a K-V and a Gaussian distribution at $\sigma_0=60^\circ$.
For the K-V distribution, no significant emittance growth is resolved above $\sigma=10^\circ$ over the simulated interval.
Below that value a small growth is obtained, consistent with the higher-order instability region of the K-V model.

The Gaussian growth factors in Fig.~\ref{fig10} are approximately bracketed by the two analytical estimates.
For the reported runs with $\sigma>10^\circ$, no significant secular growth is observed after the initial redistribution, apart from bounded oscillations.
This is consistent with homogenization as the dominant initial source of emittance growth in that range.
It is not a proof of spectral stability against all fourth- and higher-order modes, nor does it exclude small numerical contributions.

\section{CONCLUSIONS\label{sec4}}
The K-V distribution has been investigated by many authors.
For the periodic focusing systems considered in Ref.~1, it exhibits a series of instabilities associated with structure resonances.
The condition $\sigma_0\le180^\circ/n$ is the avoidance criterion quoted there for the corresponding $n$th-order structure-resonance family.
It should not be interpreted as a universal stability theorem for arbitrary distributions and focusing lattices.
Our simulations show the characteristic third-order and higher-order responses for $\sigma_0=90^\circ$.
For $\sigma_0=60^\circ$, significant long-term K-V emittance growth is absent over most of the reported range, with a small increase appearing below approximately $\sigma=10^\circ$.
The simulation studies have been extended to four non-K-V distributions: waterbag, parabolic, conical and Gaussian.
Two types of additional effects have been found:
\begin{enumerate}
 \item The visible development of the structure-resonance response and the secular emittance-growth rate generally decrease toward the Gaussian distribution in the cases studied.
 This behavior is consistent with a reduced coherent response, but the damping mechanism has not been established independently.
 \item A rapid redistribution toward a more homogeneous particle density in real space takes place within approximately the first cell.
 This leads to an initial growth of the rms emittance due to transformation of field energy into kinetic and potential beam energy.
\end{enumerate}

The non-K-V simulations exhibit distortions and emittance growth associated with the K-V instability regions.
The characteristic arms in the $x,x^\prime$ and $y,y^\prime$ projections become progressively less distinct.
For the Gaussian distribution these patterns are no longer clearly recognized, so the identity of an individual mode cannot be inferred from those projections alone.

The amount of the initial growth depends on the type of the initial distribution and also increases from the waterbag over the parabolic and conical towards the Gaussian distribution.
For each distribution we calculated a factor $f$, which defines its excess electric field energy relative to a beam of uniform spatial density at the same line charge and spatial second moments.
The Gaussian factor refers to the untruncated analytic reference.
Assigning this effective excess energy first to a rematched harmonic model and secondly to kinetic energy at fixed rms size gives the two estimates
\begin{displaymath}
 \frac{\<\varepsilon\>^+}{\<\varepsilon\>}\simeq
 \left\{\begin{array}{l}
 \displaystyle\sqrt{\left[1+f\left(\frac{\sigma_0^2}{\sigma^2}-1\right)\right]
 \left[1+f\left(1-\frac{\sigma^2}{\sigma_0^2}\right)\right]},
 \qquad\mbox{harmonic model},\\[1.2em]
 \displaystyle\sqrt{1+2f\left(\frac{\sigma_0^2}{\sigma^2}-1\right)},
 \hspace*{38mm}\mbox{fixed-size model}.
 \end{array}\right.
\end{displaymath}
These are alternative model estimates, not a strict inequality for the actual beam.
The comparison with Fig.~\ref{fig10} is consistent with the initial-redistribution interpretation for $\sigma_0=60^\circ$ in the range without significant subsequent secular growth.
In the case of $\sigma_0=90^\circ$, additional growth in the instability regions is consistent with the action of structure resonances.
The main conclusion is that the initially non-self-consistent distributions treated in this paper can exhibit rms emittance growth
through internal redistribution even when their initial first and second moments are matched to the channel.
Rms matching alone does not remove the excess energy associated with a nonuniform spatial density.

\acknowledge
This work was performed while one of us (M.~R.) was on sabbatical leave from the University of Maryland at GSI, W.~Germany.
The support from the Alexander von Humboldt Foundation and the hospitality of the GSI laboratory during this leave are gratefully acknowledged.
We would also like to thank W.~Lysenko for several helpful comments.

\end{document}